\documentclass[12pt,fleqn]{article}
\date{}

\begin{document}

\title{The creation of kappa deformed electromagnetic radiation
from a sum of zero modes}
\author 
{$^1$M. V. Cougo-Pinto{\footnote{\it marcus@if.ufrj.br}}, $^1$C. 
Farina{\footnote{\it farina@if.ufrj.br}} 
and $^{1,2}$J. F. M. Mendes{\footnote{\it jayme@if.ufrj.br}}
\\$^1${\it Instituto de F\' \i sica-UFRJ, CP 68528, Rio de Janeiro, RJ, 
21.941-972}\\$^2${\it 
IPD-CTEx, Av. das Am\'ericas 28.705, Rio de Janeiro, RJ, 23.020-470}\\}
\maketitle

\begin{abstract}
In a related paper we have obtained that the effective action for
a $\kappa$-deformed quantum field theory has a real and an imaginary part.
The real part is half the sum of the $\kappa$-deformed zero mode 
frequencies, while the imaginary part is proportional to the sum
of the squares of the zero mode frequencies, being the proportionality
constant equal to the inverse of $2\pi\kappa$. Here we calculate
this imaginary part for the $\kappa$-deformed electromagnetic field
confined between two perfectly conducting parallel plates. After
renormalization this imaginary part gives a creation rate of 
$\kappa$-deformed electromagnetic radiation. This creation rate
goes to zero at the appropriate limits, namely: when the deformation
disappears or at infinite separation of the plates. The result agrees
with previously obtained results and shed light on them by exhibiting
the creation rate as originated in a sum of zero modes. Let us note
that due to the rather complicated $\kappa$-deformed electromagnetic 
dispersion relation we were led to the theorem of the argument in order 
to sum the squares of the $\kappa$-deformed frequencies.
\end{abstract}

In a companion \cite{pc02n1} paper, we have obtained that the effective 
action of a $\kappa$-deformed field theory \cite{Lukierski91} has a real 
part that is the 
usual sum of frequencies of zero modes and an imaginary part given by 
a sum of the squares of these frequencies. For the electromagnetic field 
this imaginary part is given in terms of the parameter $q=(2\kappa)^{-1}$ 
as:
\begin{equation}
\Im\left\{\frac{\cal W}{T}\right\}=-\frac{q}{2\pi}\sum_{\bf k,\lambda}\omega_{\bf k}^2,
\label{ImW} 
\end{equation}
where $\lambda$ is the polarization of the state and $\omega_{\bf k}$
the deformed frequency\cite{pc02n1}. Let us calculate this
imaginary part for the $\kappa$-deformed electromagnetic field
confined between two perfectly conducting parallel plates of side
$\ell$ and separation $a$. Since the deformation does not
effect the spatial derivatives the boundary condition implemented by the 
conducting plates gives rise to the same spectrum of wave-numbers obtained
in the non-deformed case. We have then 
$|{\bf k}|=\sqrt{k_{\parallel}^{2}+k_{n}^{2}}$, 
where $k_\parallel$ is the modulus of the component of {\bf k} which is parallel 
to the plates, $k_n=n\pi/a$ ($n=0,1,2...$) and for all modes but $n=0$ there
is a double degeneracy due to polarization. The deformed frequencies are 
accordingly given by \cite{pc02n1}:
\begin{equation}
\omega_{\bf k}=\frac{1}{q}{\rm arcsenh}
\left(q\sqrt{k_{\parallel}^2+k_n^2}\right) \; .
\label{omegadekparaleloen} 
\end{equation}
In this way we have for the imaginary part (\ref{ImW}) of the effective action:
\begin{eqnarray}\label{ImW2a}
\Im\left\{\frac{\cal W}{T}\right\}&=&-\frac{q}{2\pi}
\frac{\ell^2}{(2\pi)^2}
\sum_{n,\sigma}\int\int
dk_1dk_2\; \omega_{\bf k}^2\nonumber\\
&=&-\frac{q}{2\pi}\frac{\ell^2}{(2\pi)^2}
\int_{0}^{\infty}2{\pi}dk_{\parallel}k_{\parallel}
\sum_{n=(0)}^{\infty}2\; 
\frac{1}{q^2}{\rm arcsenh}^2
\left(q\sqrt{k_{\parallel}^2+k_n^2}\right) \; ,
\end{eqnarray}
where the $(0)$ in the summation symbol means that the factor 2 inside in
the sum should be ignored for $n=0$. As a matter of fact the term
with $n=0$ will give rise to a spurious term in the effective action.
For the interest of simplicity we will discard it right now.
As it stands, this expression for
the imaginary part of effective action is ill defined. It must be 
regularized and set free of spurious terms to obtain its physical
meaningful part. Let us introduce an attenuation positive
parameter $\epsilon$ and define the regularized expression:
\begin{eqnarray}\label{ImW3a}
\Im\left\{\frac{\cal W}{T}\right\}&=&-\frac{q}{2\pi}
\frac{\ell^2}{(2\pi)^2}
\sum_{n,\sigma}\int\int
dk_1dk_2\; \omega_{\bf k}^2\nonumber\\
&=&-q\frac{2\ell^2}{(2\pi)^2}
\int_{0}^{\infty}dk_{\parallel}k_{\parallel}\;
S_{\epsilon}(q,k_{\parallel}) \; ,
\end{eqnarray}
where
%
%
\begin{equation}
S_{\epsilon}(q,k_{\parallel}):=
\sum_{n=1}^{\infty}
\frac{e^{-\epsilon\sqrt{k_{\parallel}^2+k_n^2}}}{q^2}{\rm arcsenh}^2
\left(q\sqrt{k_{\parallel}^2+k_n^2}\right).
\end{equation}
By taking the derivative of $q^{2}S_{\epsilon}(q,k_{\parallel})$ in 
relation to $q$ we obtain 
%
%
\begin{equation}\label{derondqdeSepsilon}
\frac{\partial}{\partial{q}}\left(q^{2}S_{\epsilon}(q,k_{\parallel})\right)
=2\sum_{n=1}^{\infty}
{\rm arcsenh}(q\sqrt{k_{\parallel}^2+k_n^2})\frac{\sqrt{k_{\parallel}^2+
k_n^2}e^{-\epsilon\sqrt{k_{\parallel}^2+k_n^2}}}{\sqrt{1+q^2(k_{\parallel}^2+k_n^2)}}
\end{equation}
We now write the following equality depending on a new parameter $\nu{\in}[0,1]$: 
\begin{equation}\label{Sepsilonenu}
\frac{\partial}{\partial{q}}\left(q^{2}S_{\epsilon,\nu}(q,k_{\parallel})\right)
=2\sum_{n=1}^{\infty}
{\rm arcsenh}(q\nu\sqrt{k_{\parallel}^2+k_n^2})\frac{\sqrt{k_{\parallel}^2+
k_n^2}e^{-\epsilon\sqrt{k_{\parallel}^2+k_n^2}}}
{\sqrt{1+q^2(k_{\parallel}^2+k_n^2)}} \; ,
\end{equation}
Taking the derivative of (\ref{Sepsilonenu}) in relation to $\nu$
we obtain:
\begin{equation}\label{derondnuderondqdeSepsilonenu}
\frac{\partial^2}{\partial\nu\partial{q}}
\left(q^{2}S_{\epsilon,\nu}(q,k_{\parallel})\right)
=2q\sum_{n=1}^{\infty}\frac{(k_{\parallel}^2+k_n^2)
e^{-\epsilon\sqrt{k_{\parallel}^2+k_n^2}}}{{\sqrt{1+q^2(k_{\parallel}^2+k_n^2)}}
{\sqrt{1+(q\nu)^2(k_{\parallel}^2+k_n^2)}}}\; .
\end{equation}
In this way, the integral of this equality in relation to $\nu$, from
$\nu=0$ to $\nu=1$ gives us back the equality (\ref{derondqdeSepsilon}).

We shall work on the sum (\ref{derondnuderondqdeSepsilonenu}) way
back to (\ref{ImW3a}) until the imaginary part of the effective action
is reduced to a series of quadratures.

To perform the sum (\ref{derondnuderondqdeSepsilonenu})
we use the principle of the argument, which states that:
\begin{equation}
\sum_{n}\phi(z_n)-\sum_{n}\phi(p_n)=\frac{1}{2{\pi}i}
\int_{\cal C}\phi(z)\frac{d}{dz}\log f(za)dz\; ,
\end{equation}
where $\phi$ is an analytic function inside and on the curve ${\cal C}$
except at a finite number of poles inside ${\cal C}$, $f$ is an analytic
function inside and on ${\cal C}$ and the sums are on the zeros $z_n$ and
poles $p_n$ of $\phi$ inside ${\cal C}$. We apply the principle to
the functions
\begin{equation}
\phi{(z)}=\frac{(k_{\parallel}^2+z^2)
e^{-\epsilon\sqrt{k_{\parallel}^2+z^2}}}{\sqrt{1+q^2(k_{\parallel}^2+z^2)}
\sqrt{1+(\nu{q})^2(k_{\parallel}^2+z^2)}}
\end{equation}
and
\begin{equation}
f(z)=\sin(az)\; .
\end{equation}
Let us notice that $\phi$ has poles at 
$z_{\pm}^\prime=\pm{i}\sqrt{k_\parallel^2+q^{-2}}$ and
$z_{\pm}^{\prime\prime}=\pm{i}\sqrt{k_\parallel^2+(q\nu)^{-2}}$,
where we should keep in mind that $0<\nu\leq{1}$. We take the branch
cuts of $\phi$ from $z_{+}^\prime$ to $z_{+}^{\prime\prime}$
and from $z_{-}^\prime$ to $z_{-}^{\prime\prime}$.
The closed curve ${\cal C}$ will consist of two parts. The first is a path 
parallel to the imaginary axis, from $iR$ to $-iR$ at a distance $\sigma$ to 
the right of the imaginary axis; $R$ is large and $\sigma$ is infinitesimal.
The second part is a anti-clockwise semicircle on the positive semi-plane,
from $-iR$ to $iR$. In this way poles of $\phi$ are outside ${\cal C}$.
By taking the limit $R\rightarrow\infty$ and $\sigma\rightarrow{0}$
we obtain for the sum (\ref{derondnuderondqdeSepsilonenu}):
\begin{eqnarray}
&&\frac{1}{2q}\frac{\partial^2}{\partial\nu\partial{q}}
\left(q^{2}S_{\epsilon,\nu}(q,k_{\parallel})\right)=\nonumber\\
&=&\frac{1}{2\pi{i}}\oint_{\cal C}\frac{(k_{\parallel}^2+z^2)
e^{-\epsilon\sqrt{k_{\parallel}^2+z^2}}}{\sqrt{1+q^2(k_{\parallel}^2+z^2)}
\sqrt{1+(\nu{q})^2(k_{\parallel}^2+z^2)}}\;
\frac{d\log{({\rm sen}(az))}}{dz}dz \; .
\end{eqnarray}
The integral on the semicircle of infinite radius is zero due to the
damping exponential $\exp\{{-\epsilon\sqrt{k_{\parallel}^2+z^2}}\}$, which
will be omitted from now on. From the integral just to the right of the
imaginary axis it only survives the parts just besides the branch cuts.
As a result we have for the sum:
\begin{eqnarray}
&&\frac{1}{2q}\frac{\partial^2}{\partial\nu\partial{q}}\left(q^{2}S_{\epsilon,\nu}
(q,k_{\parallel})\right)=\nonumber\\
&=&-\frac{1}{\pi{i}}\int_{\sqrt{k_\parallel^2+q^{-2}}}^{\sqrt{k_\parallel^2+(q\nu)^{-2}}}
\frac{k_{\parallel}^2-y^2}{\sqrt{1+q^2(k_{\parallel}^2-y^2)}
\sqrt{1+(\nu{q})^2(k_{\parallel}^2-y^2)}}
\frac{d\log{({\rm senh}(ay))}}{dy}dy.
\end{eqnarray}
The integration of both sides of this equation in $k_\parallel$ lead us to
\begin{eqnarray}
&&\frac{1}{2q}\frac{\partial^2}{\partial\nu\partial{q}}\left(q^{2}
\int_{0}^{\infty}dk_\parallel{k}_{\parallel}S_{\epsilon,\nu}(q,k_{\parallel})\right)
=\nonumber\\
&=&\frac{1}{q^2\pi\nu}\int_{0}^{\infty}dk_\parallel{k}_{\parallel}
\int_{\sqrt{k_\parallel^2+(q\nu)^{-2}}}^{\sqrt{k_\parallel^2+q^{-2}}}dy
\frac{y^2-k_{\parallel}^2}{\sqrt{k_{\parallel}^2+(q\nu)^{-2}-y^2}
\sqrt{y^2-k_{\parallel}^2-q^2}}
\frac{d\log{({\rm senh}(ay))}}{dy}\; .
\end{eqnarray}
By exchanging the order of integration in $k_\parallel$ e $y$
and performing a long and tedious calculation, we obtain: 

\[
\frac{1}{2q}\frac{\partial^2}{\partial\nu\partial{q}}\left(q^{2}
\int_{0}^{\infty}dk_\parallel{k}_{\parallel}S_{\epsilon,\nu}(q,k_{\parallel})\right)=
\]
\[
=-\frac{1}{q^2\pi\nu}\Bigg\{
\int_{(q\nu)^{-1}}^{\infty}dy\frac{d\log({\rm senh}(ay))}{dy}
\left[\frac{\pi}{4q^2}(1+\nu^{-2})\right]+
\]
\[
+\int_{q^{-1}}^{(q\nu)^{-1}}dy\frac{d\log({\rm senh}(ay))}{dy}
\Bigg[\frac{(1+\nu^{-2})}{2q^2}{\rm arcsenh}\left(\sqrt{\frac{(yq)^2-1}{\nu^{-2}-1}}\right)
\]
\begin{equation}
-\frac{1}{2q^2}\sqrt{(yq)^2-1}\sqrt{\nu^{-2}-(yq)^2}\Bigg]
\Bigg\}
\end{equation}
In these integrals there are terms proportional to the separation $a$
between the plates, since: 
\begin{equation}
\frac{d\log({\rm senh}(ay))}{dy}
=a+\frac{d\log(1-e^{-2ay})}{dy}
\end{equation}
Those terms give rise in the whole space to a uniform (and infinity) rate of 
creation of radiation per unit volume. This is a spurious contribution to
the effective action that we will discard to arrive at:
\begin{equation}
\frac{\partial^2}{\partial\nu\partial{q}}\left(q^{2}
\int_{0}^{\infty}dk_\parallel{k}_{\parallel}S_{\epsilon,\nu}(q,k_{\parallel})\right)
=\frac{2q}{\pi\nu}\int_{q^{-1}}^{(q\nu)^{-1}}dy
\frac{y^3\log(1-e^{-2ay})}{\sqrt{(yq)^2-1}\sqrt{\nu^{-2}-(yq)^2}}
\end{equation}
Now we integrate both sides of this equality in relation to $\nu$, from
$\nu=0$ to $\nu=1$, and make an exchange in the order of integration 
between $y$ and $\nu$, to arrive at: 
\begin{equation}
\frac{\partial}{\partial{q}}\left(q^{2}
\int_{0}^{\infty}dk_\parallel{k}_{\parallel}S_{\epsilon,\nu}(q,k_{\parallel})\right)=
\int_{1/q}^{\infty}dy\,y^2\frac{\log(1-e^{-2ay})}{\sqrt{(yq)^2-1}}
\end{equation}
This equation in turn is integrated from $q=0$ to arbitrary $q$ and the
order of integration is again exchanged with the integration on $y$ to
obtain: 
\begin{equation}
q^{2}\int_{0}^{\infty}dk_\parallel{k}_{\parallel}S_{\epsilon,\nu}(q,k_{\parallel})
=\int_{1/q}^{\infty}dy\,y^2\log(1-e^{-2ay})\int_{1/y}^{q}\frac{dq'}{\sqrt{(yq')^2-1}}.
\end{equation}
The integral on $q'$ can be done to recast this expression into the form:
\begin{equation}
\int_{0}^{\infty}dk_\parallel{k}_{\parallel}S_{\epsilon,\nu}(q,k_{\parallel})
=\frac{1}{q^2}\int_{1/q}^{\infty}dy\,y\log(1-e^{-2ay})\,{\rm arcosh}(yq).
\end{equation}
Now we modify this expression to arrive at an expression of the
creation rate that is easy to compare with previous results. For
this we use the series expansion 
$\log(1-e^{-2ay})=-\sum_{n=1}^{\infty}e^{-2nay}/n$ to obtain:
\[
\int_{0}^{\infty}dk_\parallel{k}_{\parallel}S_{\epsilon,\nu}(q,k_{\parallel})
=-\frac{1}{q^2}\sum_{n=1}^{\infty}\frac{1}{n}
\int_{1/q}^{\infty}dy\,y\,e^{-2nay}\,{\rm arcosh}(yq)=
\]
\[
=\frac{1}{q^2}\sum_{n=1}^{\infty}\frac{1}{2n^2}\frac{\partial}{\partial{a}}
\int_{1/q}^{\infty}dy\,e^{-2nay}\,{\rm arcosh}(yq)=
\]
\[
=\frac{1}{q^2}\sum_{n=1}^{\infty}\frac{1}{2n^2}\frac{\partial}{\partial{a}}
\left[\frac{e^{-2nay}}{-2na}{\rm arcosh}(yq)\Bigg|_{1/q}^{\infty}-
\int_{1/q}^{\infty}dy\,\frac{e^{-2nay}}{-2na}\frac{q}{\sqrt{(yq)^2-1}}\right]=
\]
\begin{equation}
=\frac{1}{q^2}\sum_{n=1}^{\infty}\frac{1}{2n^2}\frac{\partial}{\partial{a}}
\int_{1/q}^{\infty}dy\,\frac{e^{-2nay}}{2na}\frac{q}{\sqrt{(yq)^2-1}}. 
\end{equation}
Finally, we get:  
\begin{equation}
\int_{0}^{\infty}dk_\parallel{k}_{\parallel}S_{\epsilon,\nu}(q,k_{\parallel})
=-\frac{1}{2aq}\sum_{n=1}^{\infty}\frac{1}{n^2}\int_{1/q}^{\infty}dy\,
\left(y+\frac{1}{2na}\right)\frac{e^{-2nay}}{\sqrt{(yq)^2-1}}\label{10}
\end{equation}
Substituting this result in (\ref{ImW3a})
we obtain for the creation rate of radiation per unit volume:
\begin{equation}
\Im\left\{\frac{\cal W}{Ta\ell^2}\right\}=\frac{1}{(2\pi{a})^2}
\sum_{n=1}^{\infty}\frac{1}{n^2}\int_{1/q}^{\infty}dy\,
\left(y+\frac{1}{2na}\right)\frac{e^{-2nay}}{\sqrt{(yq)^2-1}} \; ,
\end{equation}
which is consistent with previous results obtained by other 
methods \cite{CougoPinto-Farina97}. Let us notice that in our final expression
it is manifest the property that the creation rate disappears when
there is no deformation ($q\rightarrow 0$) or boundary condition
($a\rightarrow\infty$).

\end{document}